\documentclass[useAMS,usenatbib]{mn2e}
\usepackage{graphicx}

\def\lesssim{\mathrel{\hbox{\rlap{\hbox{\lower4pt\hbox{$\sim$}}}\hbox{$<$}}}}
\def\gtrsim{\mathrel{\hbox{\rlap{\hbox{\lower4pt\hbox{$\sim$}}}\hbox{$>$}}}}

\def\eg{e.g.}

\title[Box Size \& Properties of Dark Matter Haloes]
      {The Impact of Box Size on the Properties of Dark Matter Haloes in Cosmological Simulations}
      \author[Power \& Knebe]
             {Chris~Power$^1$\thanks{E-mail: cpower@astro.swin.edu.au}
	       \& Alexander~Knebe$^2$\thanks{E-mail: aknebe@aip.de}\\ 
	       $^1$  Centre for Astrophysics and Supercomputing,
               Swinburne University of Technology,
               PO Box 218, Hawthorn, 3122, Victoria, Australia\\
	       $^2$ Astrophysikalisches Institut Potsdam, An der
               Sternwarte 16, 14482 Potsdam, Germany
               }
             
\date{submitted to MNRAS}

\pagerange{\pageref{firstpage}--\pageref{lastpage}}\pubyear{2006}

\begin{document}
\label{firstpage}

\maketitle

\begin{abstract}

We investigate the impact finite simulation box size has on the
structural and kinematic properties of Cold Dark Matter haloes forming in
cosmological simulations. Our approach involves generating a single 
realisation of the initial power spectrum of density perturbations and 
studying how truncation of this power spectrum on scales larger than 
$L_{\rm cut} = 2\pi/k_{\rm cut}$ affects the structure of dark matter
haloes at $z=0$. In particular, we have examined the cases of 
$L_{\rm cut} = f_{\rm cut} L_{\rm box}$ with $f_{\rm cut}=1$ (i.e. no 
truncation)$, 1/2, 1/3$ and $1/4$.
In common with previous studies, we find that the suppression of long 
wavelength perturbations reduces the strength of clustering, as
measured by a suppression of the 2-point correlation function $\xi(r)$,
and reduces the numbers of the most massive haloes, as reflected in the 
depletion of the high mass end of the mass function $n(M)$.
Interestingly, we find that truncation has little impact on the internal 
properties of haloes. The masses of high mass haloes decrease in a 
systematic manner as $L_{\rm cut}$ is reduced, but the distribution of
 concentrations is unaffected. On the other hand, the 
median spin parameter is $\sim 50\%$ lower in runs with $f_{\rm cut}<1$. 
We argue that this is an imprint of the linear growth phase of the halo's 
angular momentum by tidal torquing, and that the absence of any measurable 
trend in concentration and the weak trend observed in halo shape
reflect the importance of virialisation and complex mass accretion
histories for these quantities. These results are of interest for studies 
that require high mass resolution and statistical samples of simulated
haloes, such as simulations of the population of first stars. Our
analysis shows that large-scale tidal fields have relatively little
effect on the internal properties of Cold Dark Matter haloes and hence
may be ignored in such studies.

\end{abstract}

\begin{keywords}
cosmology:theory - dark matter - gravitation
\end{keywords}

\setcounter{footnote}{1}


\section{Introduction}
\label{sec:intro}

Over the last twenty years, the cosmological N-Body simulation 
has come to be firmly established as the preeminent tool for theoretical 
studies of the formation and evolution of structure in the Universe 
(see \citet{bert98} for a relatively recent review). Whereas early
studies explored the gross features of different cosmological 
models and led to the emergence of the Cold Dark Matter (CDM) paradigm
\citep{frenk83,white83,davis85}, the most recent generation of
simulations are being used to provide detailed predictions of the CDM 
model and in particular its $\Lambda$CDM variant. These include the behaviour 
of CDM halo mass profiles on scales that can be directly compared with 
observations
\citep[e.g.][]{hayashi04,navarro04,diemand04} and sophisticated 
semi-analytical models of galaxy formation built upon merger trees harvested 
from high resolution simulations \citep[c.f. the Millennium Simulation 
described in][]{springel05}. The N-Body approach is not without its 
limitations, however, and given its significance, it is of fundamental 
importance to understand what these limitations are and how they affect 
the interpretation of the results of simulations. 

Such considerations generally lead to ``convergence'' studies that seek to 
quantify the degree to which the results of a simulation are affected by 
choices made in its setting up and running. For example, both \citet{nfw96} 
and \citet{moore98} demonstrated that their results on the form of CDM halo 
mass profiles were not compromised by force resolution (or ``softening'') as 
well as starting redshift \citep[][]{nfw96} and particle number 
\citep[][]{moore98}. The recent study of \citet{power03} established a set of 
convergence criteria \citep[reaffirmed by][]{hayashi04} that determine the 
radial range over which the circular velocity profile of a simulated CDM halo
 is reliable to better than $10\%$ accuracy, for a number of simulation 
parameters including softening, time integration accuracy, and particle 
number \citep[see also][]{diemand04,binney04}. More generally, \citet{alex} 
investigated how properties such as the correlation function, mass function 
and central densities of dark matter haloes varied with softening and time 
integration accuracy at fixed particle mass in simulations run with different 
N-Body codes.\\

Recently, \citet[][hereafter BR05]{bagla05} have demonstrated that size
of simulation box $L_{\rm box}$ can also affect certain properties of
the dark matter halo population, in particular the 2-point correlation 
function $\xi(r)$ and the mass function $n(>M)$. $L_{\rm box}$ sets the 
longest wavelength perturbation that can be resolved in a simulation and 
it follows that reducing $L_{\rm box}$ suppresses the contribution of 
large scale perturbations to the power spectrum. The effect of this
suppression was highlighted by \citet[][hereafter GB94]{gelb94b}, who 
examined the importance of long wavelength perturbations in the initial
conditions for $\xi(r)$ and the correlation length $r_0$ in the context
of the SCDM model. They investigated how $\sigma_8$ (the linear mass 
variance in a sphere of radius $8 h^{-1}\,{\rm Mpc}$) and consequently 
$\xi(r)$ varied with $L_{\rm box}$ (see their figures 1 and 2), and
found that $L_{\rm box} \gtrsim 100 {\rm Mpc}$ (or $50 h^{-1} {\rm
  Mpc}$ for $h=0.5$) was required to correctly recover both $\sigma_8$
and the amplitude of $\xi(r)$. In other words, studies that wish
to accurately characterise the clustering properties of dark matter 
haloes require large simulation boxes.

How large the simulation box must be to accurately recover $\xi(r)$ was 
the question addressed by BR05. They noted that finite box size can
affect not only $\xi(r)$ but also the high mass end of the mass
function $n(>M)$; as $L_{\rm box}$ is reduced below some threshold
value, the numbers of the most massive haloes decrease in a systematic
fashion. The mass function $n(>M)$ can be characterised by the
\citet{ps74} form which is a relatively simple function of
the linear mass variance $\sigma(M)$. For a given mass
$M = {4\pi/3} \bar{\rho} R^3$ with $\bar{\rho}$ mean density,
\begin{equation}
\label{eq:sigma}
{\sigma^2(M) = \int^\infty_{2\pi/{L_{\rm box}}} P(k) W^2(kR) d^3k,}
\end{equation}
where $P(k)$ is the linear power spectrum, $W^2(kR)$ is the top-hat 
filter, and the lower limit of the integral $2\pi/L_{\rm box}$ 
corresponds to the fundamental mode in the simulation box. BR05
investigated how varying $L_{\rm box}$ impacted on the form of $n(>M)$
and used this as a simple criterion for determining how large $L_{\rm box}$
must be to recover $n(>M)$ (and consequently $\xi(r)$) to a given
accuracy.\\

The studies of GB94 and BR05 clearly indicate that large simulation
boxes are required if we wish to recover accurate mass and two point 
correlation functions of the dark matter halo population, but does
the choice of $L_{\rm box}$ also affect the internal properties of the 
haloes? This question is of interest for studies that require the high 
spatial resolution offered by simulations of small volumes but that may
not require accurate clustering information, such as studies of the
first objects at high redshift \citep[e.g.][]{abel02}, and it is one
that we shall address in the present study.

Why might we expect finite $L_{\rm box}$ to be of importance for the
internal structure of haloes? The suppression of long wavelength
perturbations will modify the ``global'' tidal field that a dark
matter halo is subject to, which may have an impact on the halo's
shape, its angular momentum content \citep[e.g.][]{white84} and the
infall pattern of substructures \citep[e.g.][]{aubert04,benson05}. 
Similarly, the linear mass variance controls the formation time of dark
 matter haloes \citep[][]{nfw96,bullock01,ens01} and so we might expect 
haloes to form at systematically later times in smaller boxes with
lower ``effective'' values of $\sigma_8$; this may then affect both the 
mass of the halo and its central density (or concentration) measured at $z=0$. 
We also note that small- and large-scale modes are coupled during the 
non-linear clustering phase and there is a transfer of power from 
large scales down to small scales \citep[\eg][]{bagla97}; neglecting 
this power may leave an imprint on the formation and evolution of 
gravitationally bound objects. On the other hand, dark matter haloes are, by 
definition, virialised systems and much of the information that was 
present during the initial stages of their collapse will be erased 
during virialisation \citep[c.f. the universal mass 
profile of][]{nfw96,nfw97}, so it is not clear whether a finite $L_{\rm
  box}$ will have an effect at all. Nevertheless, it is important to
investigate this question and determine how severe a limitation it
might be.\\

The outline of this paper is as follows; in the next section, we
briefly describe the simulations we have used in this study before 
presenting our results (\S~\ref{sec:results}). We confirm the findings
of GB94 and BR05 (\S~\ref{ssec:bulk}) before examining how various
characteristic properties of the halo population -- concentrations,
shapes and spins -- are affected by the suppression of long wavelength
perturbations on scales larger than $L_{\rm cut} = f_{\rm cut}\,L_{\rm
  box}$. Finally, we discuss our results in \S~\ref{sec:discussion} and
offering some concluding remarks in \S~\ref{sec:conclusions}.

\section{The Simulations}
\label{sec:simulations}

We have run a set of cosmological N-body simulations following the
formation and evolution of structure in a set of simulation boxes of
side $L_{\rm box}=128 h^{-1} {\rm Mpc}$, assuming a flat cosmology with 
$\Omega_0=0.333$, $\Omega_{\Lambda}=0.667$, $h=0.667$ and
$\sigma_8=0.88$. Each simulation is single-mass and contains $256^3$ 
particles, implying a particle mass of $m_p=1.15\times10^{10} 
h^{-1}{\rm M_{\odot}}$; we adopt a starting redshift of $z_{\rm
  start}=40.0$ in each case. For each of the runs, we use the same 
realisation (i.e. the same amplitudes and phases) of the power 
spectrum\footnote{Generated using CMBFAST \citep[][]{cmbfast}} but we 
truncate the power on scales $k \le k_{\rm cut}=2\pi/L_{\rm cut}$, 
where $L_{\rm cut}$ is a fraction $f_{\rm cut}$ of $L_{\rm box}$ and 
$f_{\rm cut} = 1$ (i.e. no truncation), $1/2, 1/3$ and $1/4$. These 
correspond to cut-offs of $L_{\rm cut} = 128, 64, 43$ and 
$32\,h^{-1}\,{\rm Mpc}$ or $k_{\rm cut} \simeq 0.05, 0.1, 0.15$ and 
$0.2\,h\,{\rm Mpc}^{-1}$ respectively. Truncating the power spectrum in 
this way mimics the effect of varying $L_{\rm box}$ while allowing the 
impact of the truncation on individual haloes to be investigated. 

The simulations were performed using the parallel TreePM code
\texttt{GADGET2} \citep[][]{springel05} with a fixed comoving 
softening of $0.01 h^{-1} {\rm Mpc}$ and a timestep parameter of
$\eta=0.02$. 

Groups were identified using \texttt{AHF}\footnote{\texttt{AHF} can be 
downloaded from the following web page 
\texttt{http://www.aip.de/People/aknebe/AMIGA}}, a modification of the 
\texttt{MHF} algorithm that was presented in \citet{gill04}. Haloes are 
located as peaks in an adaptively smoothed density field of the
simulation using \texttt{MLAPM}'s grid hierarchy and a refinement
criterion that matches the force resolution of the actual simulation
carried out with \texttt{GADGET2} (i.e. 5 particles per cell); local
potential minima are computed for each of these peaks and the set of
particles that are gravitationally bound to the halo are returned. For
every halo (either host or satellite) we calculate a suite of canonical 
properties (e.g. velocity, mass, spin, shape, concentration, etc.)
based upon the particles within the virial radius. The virial radius 
$R_{\rm vir}$ is defined as the point where the density profile
(measured in terms of the cosmological background density $\rho_b$)
drops below the virial overdensity $\Delta_{\rm vir}$, i.e. 
$M(<R_{\rm vir})/(4\pi R_{\rm vir}^3/3) = \Delta_{\rm vir}
\rho_b$. This threshold $\Delta_{\rm vir}$ is based upon the
dissipationless spherical top-hat collapse model and is a function of
both cosmological model and time. For $z=0$ it amounts to 
$\Delta_{\rm vir}=340$.

\section{Results}
\label{sec:results} 

\begin{figure*}
  \centering
  \includegraphics[width=16cm]{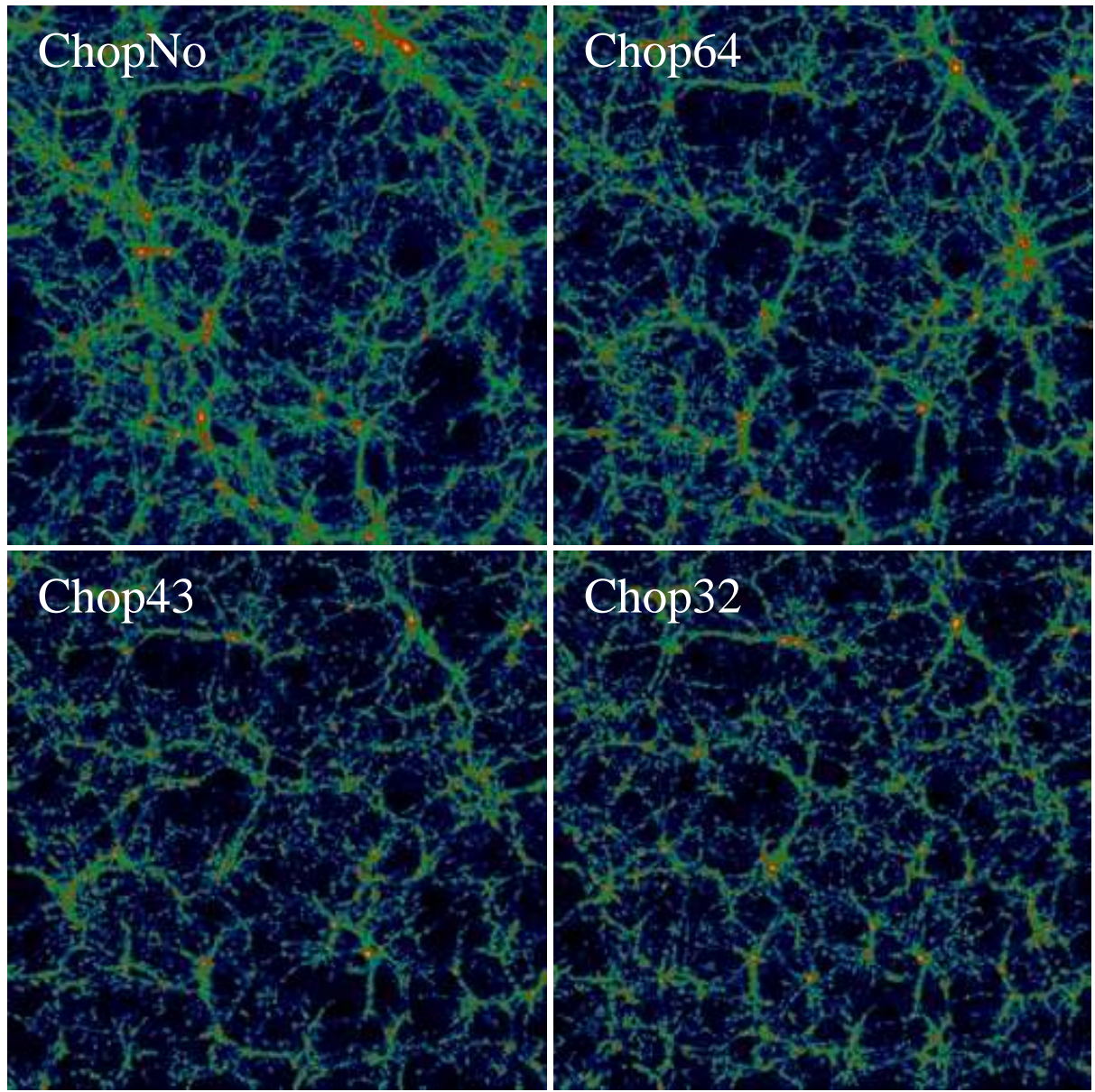}
  \caption{The projected density distribution in $12.8 h^{-1} \rm Mpc$
  slices taken through the centres of each of the boxes. We have
  smoothed the particle mass using an adaptive Gaussian kernel and
  projected onto a mesh. Each mesh point is weighted according to the
  logarithm of its projected surface density, and so the ``brighter'' the
  mesh point, the higher the projected surface density.}
\label{fig:slices}
\end{figure*}

We now present the results of our analysis. Firstly, we give a visual 
impression of the impact of truncating the power spectrum at large
scales on the overall clustering pattern within the simulation
volume. Secondly, we consider some bulk measures of the dark matter
halo population, including the mass function $n(>M)$, 2-point
correlation function $\xi(r)$ and power spectrum $P(k)$.  Finally we 
examine the internal properties of haloes, including concentrations,
shapes and spin parameters.

In what follows, we compare and contrast the runs in which the initial
power spectrum is truncated with the fiducial run \texttt{ChopNo},
whose initial power spectrum was not truncated (i.e. $f_{\rm cut}=1$). 
The ``truncated'' runs are called \texttt{Chop64} (i.e. 
$f_{\rm cut}=1/2$), \texttt{Chop43} (i.e. $f_{\rm cut}=1/3$) and
\texttt{Chop32} (i.e. $f_{\rm cut}=1/4$), respectively.

\subsection{Spatial Distribution at $z=0$ : A Visual Impression}
\label{ssec:visual}

Figure~\ref{fig:slices} shows projected maps of the dark matter density 
distribution within a thin slice taken through each of the four
simulations at $z=0$; for clarity, we have chosen a slice thickness of 
$10\%$ the box length (i.e. $12.8 h^{-1} \rm Mpc$) taken through the 
centre of the box. 

Although it is possible to pick out features that are common to all 
four runs, the magnitudes of the corresponding overdensities decrease
as $L_{\rm cut}$ is decreased. The network of filaments in the fiducial 
run is more striking than in the ``truncated'' runs -- that is, 
the fraction of volume that is at or above a given overdensity threshold 
is smaller in the runs in which large scale power was suppressed 
The central overdensities of individual haloes are comparable in the 
respective runs, but these haloes tend to occupy regions that are less
overdense in the ``truncated'' runs. Similarly, the local mean number density
of haloes appears lower in the ``truncated'' runs. Both of these latter
observations are consistent with the GB94 and BR05 results that finite
box size impacts on $\xi(r)$ and consequently the correlation length $r_0$.

\subsection{Bulk Properties of the Halo Population}
\label{ssec:bulk}
The projected maps of the dark matter density distribution suggest that 
haloes are less clustered and that there are fewer massive haloes in
the ``truncated'' runs. We now quantify these differences by
studying the halo mass function $n(>M)$, the correlation function
$\xi(r)$ and the non-linear power spectrum $P(k)$.\\

\begin{figure}
  \centering
  \includegraphics[width=8cm]{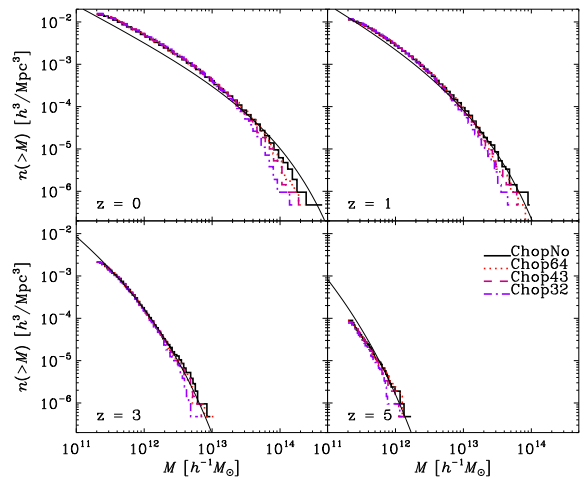}
  \caption{\textbf{The Cumulative Mass Function $n(>M)$.} We show the
  number of haloes with mass greater than $M$ in each of the runs at
  redshifts $z=5, 3, 1$ and $0$. The heavy solid curves correspond to
  the fiducial run while the dotted, dashed and dotted-dashed curves
  correspond to the runs set up with truncated power spectra. We also
  plot the predicted Press-Schechter mass function for the appropriate
  redshift (light solid curves). See text for further details.}
  \label{fig:massfunc}
\end{figure}

\begin{figure}
  \centering
  \includegraphics[width=8cm]{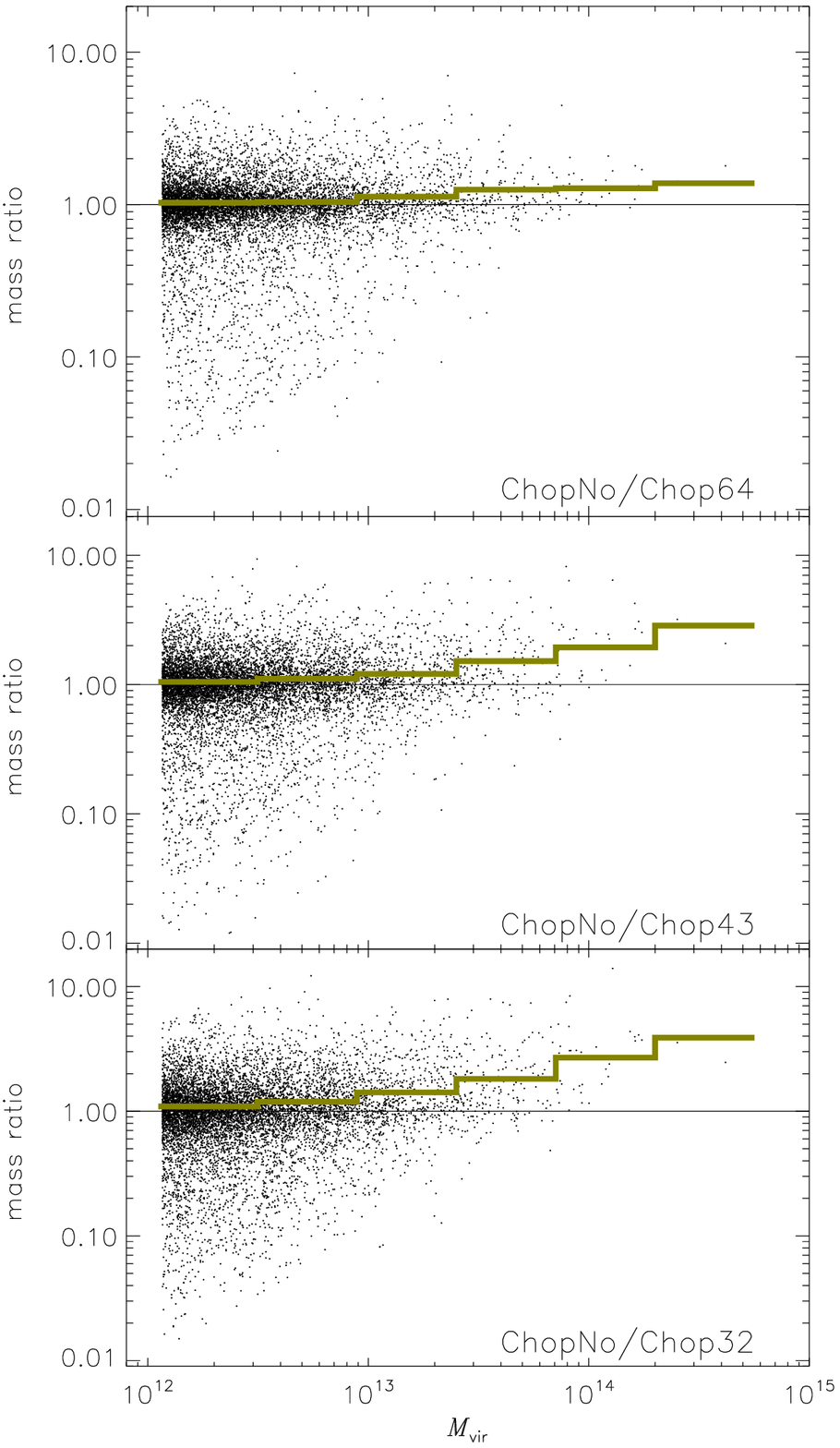}
  \caption{Comparison of Halo Masses in the truncated and fiducial runs.
   See text for further details.}
  \label{fig:MassMass}
\end{figure}

We begin by comparing cumulative halo mass functions $n(>M)$ for
each of the runs in figure~\ref{fig:massfunc}; these mass functions
have been constructed from the \texttt{AHF} halo catalogues at
redshifts $z=0, 1, 3$ and $5$. The heavy solid curves correspond to the 
fiducial \texttt{ChopNo} run while the dotted, dashed and dotted-dashed
curves correspond to the \texttt{Chop64}, \texttt{Chop43} and 
\texttt{Chop32} runs respectively. For reference, we also plot
the appropriate Press-Schechter mass function for the given redshift
(light solid curve in each of the panels).

This figure demonstrates that truncating the power on large spatial
scales leads to a suppression of the number of high mass systems that
form, and the effect is most noticeable in the runs where the
truncation is most severe. We note that the effect is most pronounced 
at $z=0$ but it is apparent at all redshifts.

We investigate this effect further in figure~\ref{fig:MassMass}, where we
cross correlate the masses of individual haloes that have been
matched in the fiducial and truncated runs, i.e. we plot the mass ratio
of corresponding haloes against the mass in the fiducial run. This is
achieved by identifying its $10\%$ most bound particles in the
\texttt{ChopNo} model and locating those haloes in the truncated runs
that contain the largest fraction of these particles; this allows for
unique matching.  

Here we observe a clear trend -- the masses of the high mass systems 
are systematically smaller in the truncated runs relative to the 
fiducial run, and the effect becomes more pronounced as the truncation 
is applied more aggressively (compare the upper and lower panels). However,
 as we go to lower masses ($10^{12} - 10^{13} h^{-1}\,{\rm M_{\odot}}$), we
find that the median halo mass is unaffected, although we note that the mass 
of any given halo can change by as much as a factor of 10 (with a rms variation
of a factor of 2).\\

\begin{figure}
  \centering
  \includegraphics[width=8cm]{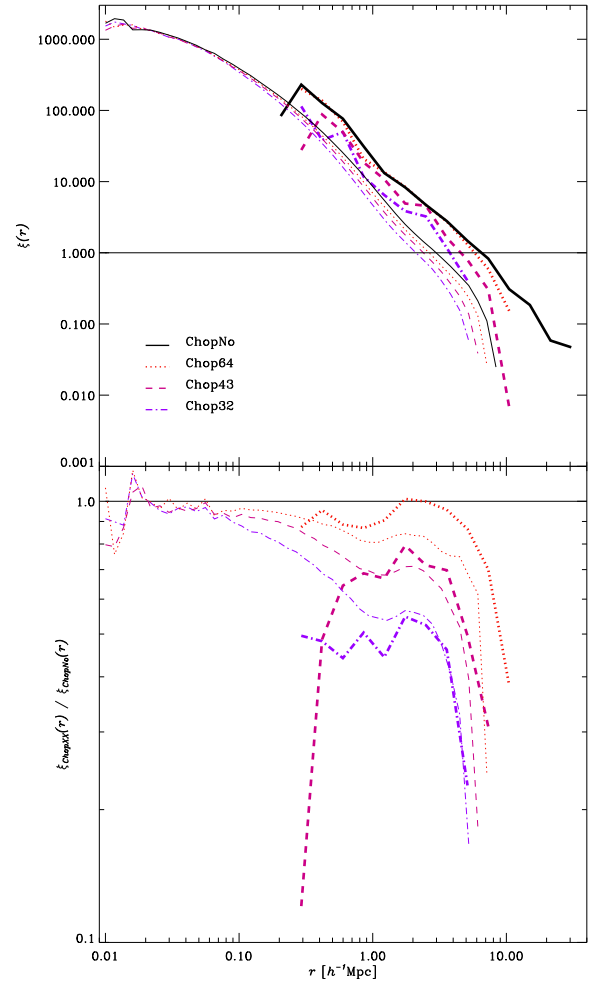}
  \caption{The 2-Point Correlation Function $\xi(r)$ (upper panel) and the 
   ratio of $\xi_{\rm ChopXX}(r)$ in the truncated runs relative to the 
   fiducial run $\xi_{\rm ChopNo}(r)$ (lower panel) for the dark matter
   (light curves) and the haloes (heavy curves).}
  \label{fig:xi}
\end{figure}

Figure~\ref{fig:xi} shows how clustering as measured by the correlation
function $\xi(r)$ varies between different runs. The heavy lines correspond to
the 850 most massive haloes within the simulation volume (according to
a number density of roughly $4\times10^{-4} h^3 {\rm Mpc}^{-3}$), while the
light lines correspond to the dark matter. Solid, dotted, dashed, 
dotted-dashed curves represent the fiducial runs and the truncated runs 
respectively. The upper panel shows $\xi(r)$ while the lower panel
shows the ratio of $\xi(r)$ measured for each of the truncated runs 
relative to the fiducial run. The first point to note is that the
haloes are more clustered than the dark matter -- a well-known result. The
second point is that the amplitude of $\xi(r)$ for both the dark matter
and the haloes decreases as the cut-off scale $L_{\rm cut}$
decreases; the extent of the effect is clearly illustrated in the
lower panel.\\

\begin{figure}
  \centering
  \includegraphics[width=8cm]{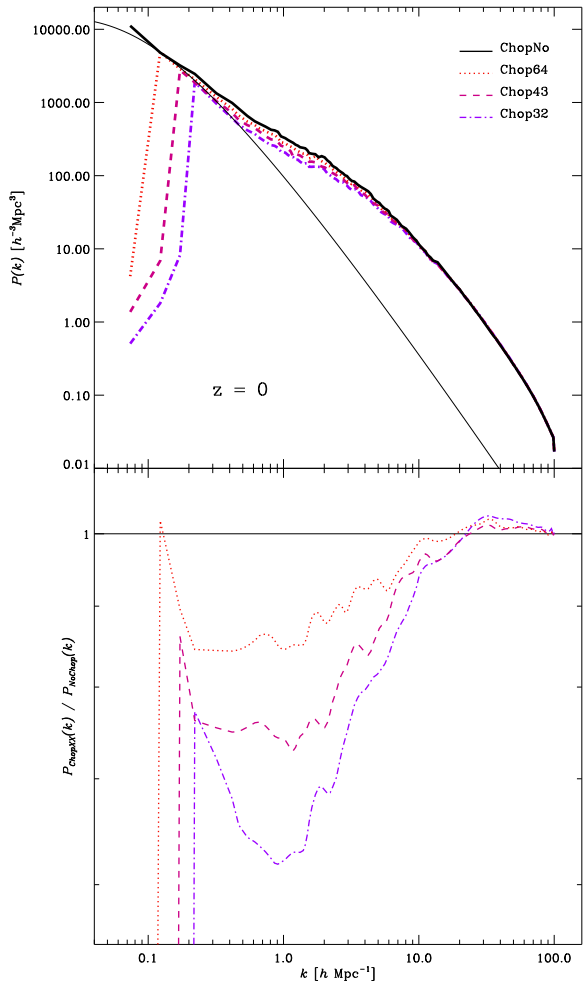}
  \caption{The Dark Matter Power Spectrum P(k) (upper panel) and the ratio of 
   power in a given mode in the truncated runs relative to the fiducial run 
   (lower panel).}
  \label{fig:pk}
\end{figure}

As a final measure of the impact of truncating the initial power
spectrum, we measure the non-linear dark matter power spectrum at $z=0$
for each of the runs, shown in figure~\ref{fig:pk}. At small scales
there is good agreement but deviations from the fiducial run become
apparent at $k \simeq 10 h \rm Mpc^{-1}$, at which point the amplitude
of the power spectrum is suppressed in the truncated runs relative to
the fiducial run. This is clear in the lower panel where we show the
ratio of the power spectra to the fiducial case.

\subsection{The Internal Properties of the Halo Population}
\label{ssec:internal}

We have so far considered \emph{bulk} properties of the halo population
and shown that our results are consistent with those of GB94 and BR05.
However, our primary motivation for this study is to determine whether 
the choice of box size influences the structural and kinematic
properties of haloes, and this is the subject of this section.
To this extent we are focusing on three relevant integral properties 
of dark matter haloes, namely concentration, spin parameter, and triaxiality.\\

We begin by investigating the impact of the finite box size on
concentration, which provides a measure of the central density 
of a halo. Typically a concentration $c_{\rm vir}$ is defined;
\begin{equation}
{c_{\rm vir}=\frac{r_{\rm vir}}{r_s}}
\end{equation}
where $r_s$ is a ``scale'' radius \citep[c.f.][defined as the radius at 
which the logarithmic slope of the density profile is $-2$]{nfw97}.
In this study, we have adopted $C_{1/5}$ \citep[\eg][]{Colin00,Knebe02}, 
defined as
\begin{equation}
{ C_{1/5} = \frac{R(1/5\,M_{\rm vir})}{R_{\rm vir}}.}
\end{equation}
$R(1/5\,M_{\rm vir})$ is the radius enclosing $1/5$ the virial
mass of the halo. This is a convenient, robust and model-independent 
measure of concentration, although it is a non-trivial function of 
$c_{\rm vir}$; for example, $c_{\rm vir} = 5$ corresponds to 
$C_{1/5} = 5$ while $c_{\rm vir} = 10$ is approximately $C_{1/5} = 7$. 
However, this is unimportant because we are interested in relative 
variations in concentration. 

We argued in \S~\ref{sec:intro} that the suppression of long wavelength
perturbations affects the linear mass variance $\sigma(M)$, which
should affect \emph{when} the halo forms and hence what its
concentration at $z=0$ will be. Furthermore, previous studies have shown that 
$c_{\rm vir}$ is a function of halo mass \citep{nfw97,ens01,bullock01} and 
the density of the environment in which the halo resides
\citep[][hereafter B01]{bullock01}. It follows that $c_{\rm vir}$ might be 
influenced by the finite box size;

\begin{itemize} 

\item The relation between halo mass and $c_{\rm vir}$ is
  such that $c_{\rm vir}$ is a decreasing function of increasing halo 
  mass. Higher mass systems tend to form at later times when the mean 
  density of the Universe was lower, and so the central densities and hence
  concentrations of these haloes are smaller. Suppressing the contribution 
  from long wavelength perturbations results in a lower effective $\sigma_8$,
  which should delay the formation time of haloes and hence reduce $c_{\rm
  vir}$. We expect this effect to be more dramatic for higher mass systems
  (c.f. figure~\ref{fig:MassMass}).
  
\item Haloes in less dense environments tend to have lower
  concentrations than haloes of similar mass in higher density
  environments (B01). Figure~\ref{fig:xi} shows that both
  the dark matter and haloes are less clustered in the truncated runs and
  thus reside in lower density environments relative to their counterparts in
  the fiducial run. 

\end{itemize}

Although there are two competing effects, we note that B01
observed that the trend with environment is stronger than the trend with mass. 
Moreover, it is the masses of the higher mass systems ($\gtrsim 10^{13} 
h^{-1}\,{\rm M_{\odot}}$) that are most affected, where the effect on 
clustering affects both the dark matter and the haloes in a similar manner.
Based on these observations, we might expect a shift towards higher 
concentrations.

\begin{figure}
  \centering
  \includegraphics[width=8cm]{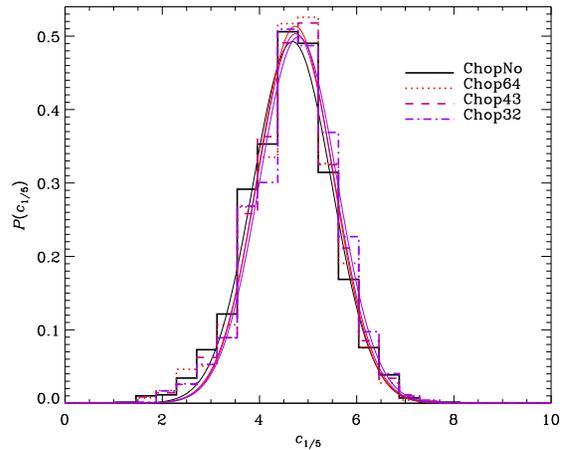}
  \caption{Distribution of Concentrations $C_{1/5}$. Solid,
  dotted, dashed and dotted-dashed histograms correspond to the
  \texttt{ChopNo} (fiducial),\texttt{Chop64}, \texttt{Chop43}
  and \texttt{Chop32} runs respectively. The thin solid lines are derived by fitting the distribution to a Gaussian.}
  \label{fig:Pconc}
\end{figure}

\begin{figure}
  \centering
  \includegraphics[width=8cm]{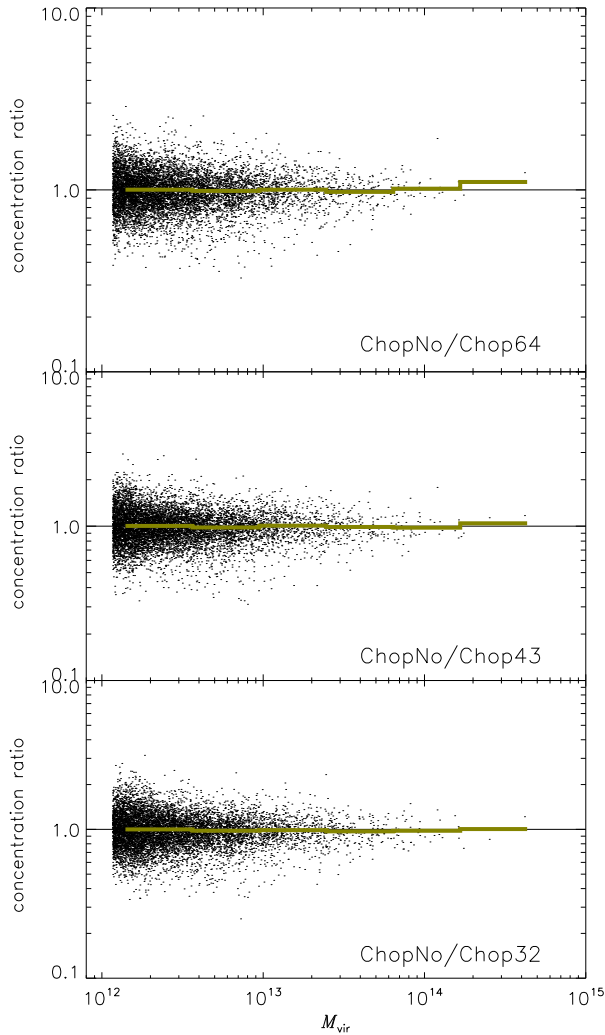}
  \caption{Comparison of halo concentrations $C_{1/5}$ in the truncated and fiducial runs.}
  \label{fig:ConcConc}
\end{figure}

In figure~\ref{fig:Pconc}, we show the distributions of $C_{1/5}$ for
the four runs. Each peaks at $C_{1/5} \simeq 5$ ($c_{\rm vir} \simeq
5$), and while there are small differences in amplitude and offset in the
position of the peak, there is no indication that the differences
between the distributions are significant. This is confirmed by computing
 the KS statistic for the truncated runs compared with the fiducial run; the
probability that the differences are significant is small ($\lesssim 10\%$) 
and negligible compared to the variation between runs with different 
realisations of the (untruncated) power spectrum. Restricting our analysis to
only those haloes more massive than $10^{13} h^{-1}\,{\rm M_{\odot}}$ does
not affect the result. 

We checked for differences in both the mass accretion histories and the mean
local density of haloes (c.f. figure 6 of B01) between the
truncated and fiducial runs, but again the differences are negligible.
Haloes in each of the runs accrete similar fractions of their final mass 
between $z=1$ and $z=0$ (e.g. $\sim 50\%$ for $10^{12} 
h^{-1}\,{\rm M_{\odot}}$ haloes) and the mean local density with a sphere of
$1 h^{-1} {\rm Mpc}$ centred on each halo in the different runs correlates 
with $C_{1/5}$ as we would expect. It seems that the concentration of 
individual haloes are sensitive to the suppression of long wavelength 
perturbations in the initial conditions (as we shall see in 
\S~\ref{ssec:case_study}), but those of a ``typical'' halo do not.
This result is further confirmed when plotting the analogue to figure~\ref{fig:MassMass}, i.e. the cross-correlation between the individual halo concentrations as shown in figure~\ref{fig:ConcConc}. We again observe a scatter about the mean ratio, but on average this ratio is found to be unity.\\

\begin{figure}
  \centering
  \includegraphics[width=8cm]{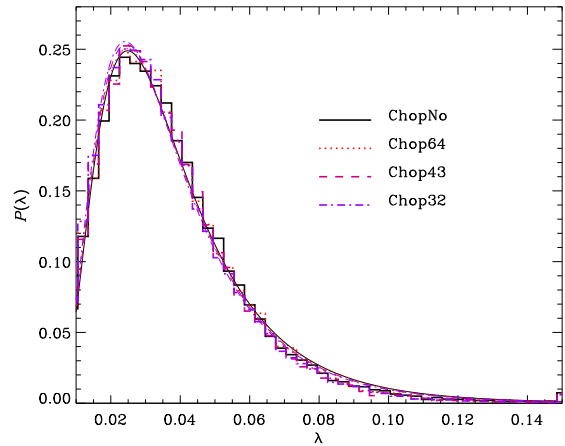}
  \caption{Distribution of Spin Parameters in the fiducial (solid curve and 
   histogram) and truncated (dotted, dashed and dotted-dashed curves and
   histograms) runs. Curves represent fits to the lognormal distribution 
   favoured by \citet{bullock01a}.}
  \label{fig:SpinFit}
\end{figure}

We have concentrated on the mass profile so far, but we also argued that the 
angular momenta and the shapes of haloes should be affected by the suppression
of long wavelength perturbations in the initial conditions. 
Figure~\ref{fig:SpinFit} shows the distribution of spin parameters $\lambda'$
in the truncated and fiducial runs. Note that we have adopted the
\citet[][hereafter B01a]{bullock01a} definition in this figure;
\begin{equation}
{\lambda' = \frac{J}{\sqrt{2} MVR}
 .}
\end{equation}
The light curves correspond to fits to the lognormal distribution advocated by
B01a\footnote{We confirm that these distributions are in good agreement with those derived for spins defined by $\lambda = J |E|^{1/2}/G M^{5/2}$.}.

\begin{figure}
  \centering
  \includegraphics[width=8cm]{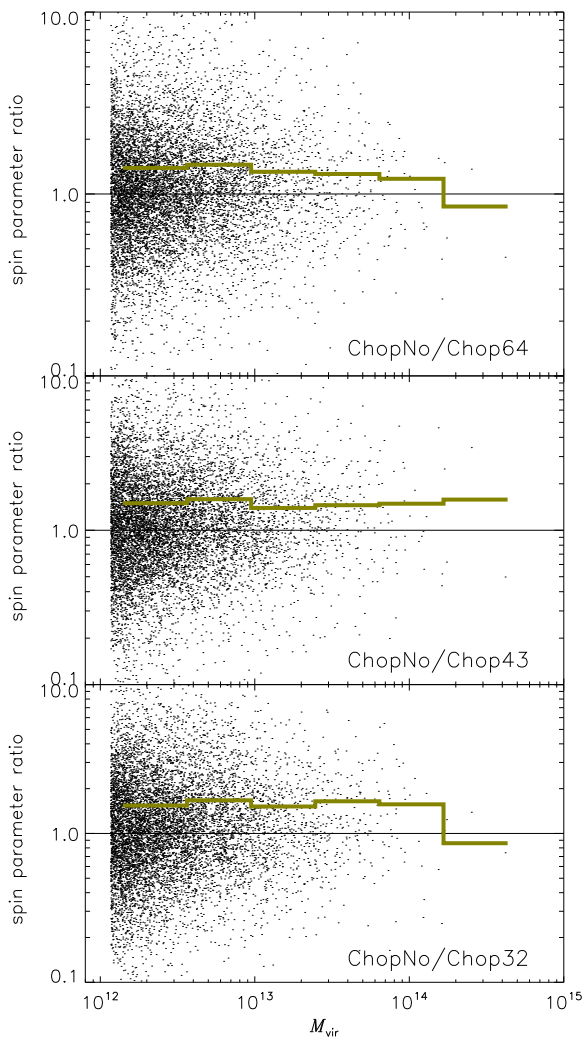}
  \caption{Comparison of Halo Spin Parameters in the fiducial and truncated 
   runs. See text for further details.}
  \label{fig:LamLam}
\end{figure}

The shapes and amplitudes of the distributions are all well described by the B01a lognormal functional form. And as the truncated models only display a marginal shift towards smaller spin parameters in figure~\ref{fig:SpinFit}, the effects of the (missing) large-scale modes are better revealed in figure~\ref{fig:LamLam}, which can be considered an analogue to 
figure~\ref{fig:MassMass} again; we match haloes in the truncated runs with their
counterparts in the fiducial run that share the largest fraction of the $10\%$
most bound particles of the halo. We can now see that there is an offset of $\sim 
50 \%$ between the fiducial run and the truncated runs, independent of halo
mass; that is, the median spin is $\sim 50\%$ larger in the fiducial run 
for haloes of all masses. We also note that the spin of an individual halo
 can vary by as much as a factor of 10 (rms variation of a factor of $\sim 3$)
between truncated and fiducial runs.\\

\begin{figure}
  \centering
  \includegraphics[width=8cm]{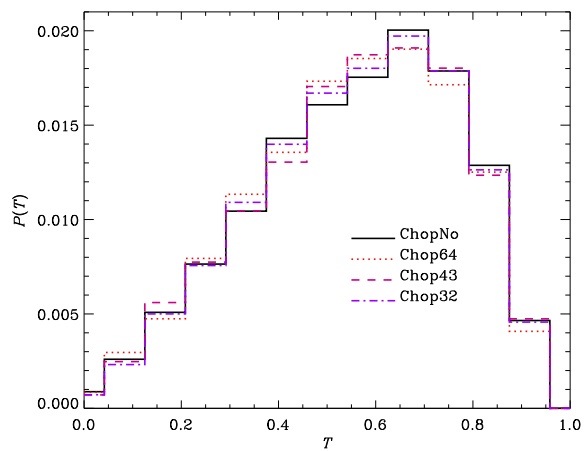}
  \caption{Distribution of Triaxialities in the fiducial and truncated runs.
  See text for further details.}
  \label{fig:PT}
\end{figure}

\begin{figure}
  \centering
  \includegraphics[width=8cm]{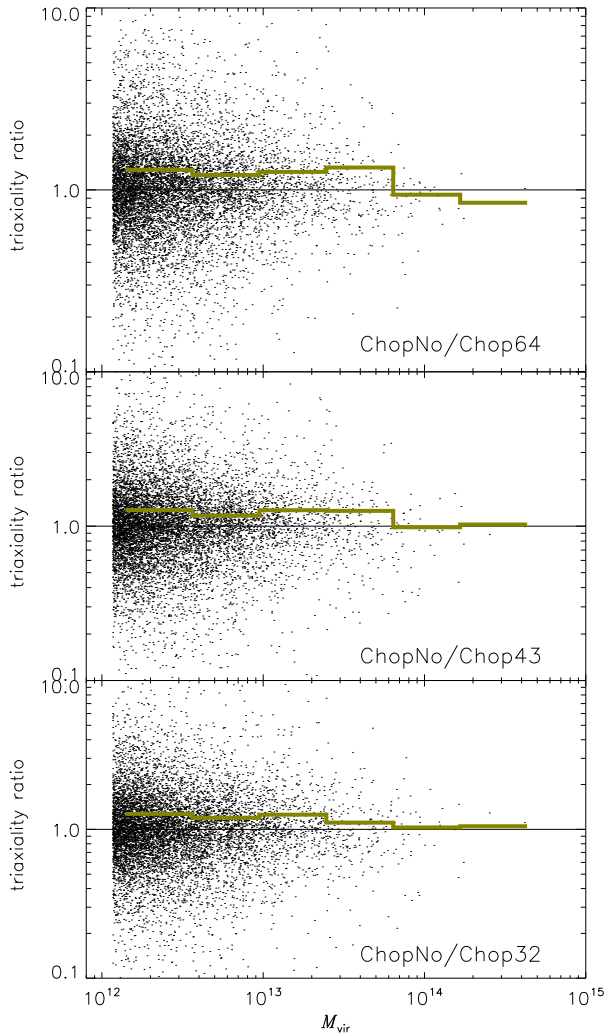}
  \caption{The ratio of the triaxiality parameter $T$ in cross-identified halos.}
  \label{fig:TT}
\end{figure}

Finally, we examine the shapes of haloes forming in the truncated runs
and compare with those that form in the fiducial run. For convenience,
we compute the \emph{triaxiality} parameter that was introduced by
\citet{franx91}, defined as
\begin{equation}
\label{eq:triaxiality}
{T = \frac{a^2-c^2}{a^2-b^2}} \ ,
\end{equation}
where $a \ge b \ge c$ are the lengths of the major, intermediate and
minor axes, respectively, as defined by all particles interior to $R_{\rm vir}$. The resulting distributions are shown in figure~\ref{fig:PT}. We observe again a trend for a marginal decrease in triaxiality when reducing the power on large scales. This is further confirmed by figure~\ref{fig:TT} where we plot the ratio of triaxiality for all cross-identified halos. The signal is similar (yet weaker) to the one already reported for the spin parameter: the more large-scale power has been ignored the smaller the triaxiality. However, it is not as pronounced, especially not for the ratio of \texttt{ChopNo} to \texttt{Chop32}.

\subsection{Case Study}
\label{ssec:case_study}

\begin{table*}
\begin{center}
\caption{\textbf{Physical Properties of Halo}. We show the virial mass,
  $M_{\rm vir}$; the virial radius, $R_{\rm vir}$; the radius at which
  the circular velocity peaks, $R_{\rm max}$; the peak circular
  velocity, $V_{\rm c,max}$; the NFW concentration, $c_{\rm vir}$; the
  spin parameter, $\lambda$; and the axis ratios, $b/a$ and $c/a$.}

\vspace*{0.3 cm}

\begin{tabular}{lcccccccc}
\hline
$f_{\rm cut}$  & $\rm M_{\rm vir}$ & $\rm R_{\rm vir}$ & $\rm R_{\rm max}$ &
        $\rm V_{\rm c,max}$ & $\rm c_{\rm vir}$ & $\lambda$ &
        ($b/a$,$c/a$) & T \\

        & [$10^{14} h^{-1} {\rm M_{\odot}}]$ & [$h^{-1} {\rm Mpc}$] &
        [$h^{-1} {\rm Mpc}$] & $[{\rm kms^{-1}}]$ & & & & \\

\hline

1         & 5.336 & 1.65 & 0.32 & 1384 & 8.3 & 0.0089 & (0.73,0.43) & 0.57  \\

1/2 & 2.625 & 1.30 & 0.35 & 1023 & 6.3 & 0.0132 & (0.63,0.42) & 0.73  \\
 
1/3 & 2.475 & 1.28 & 0.41 & 1033 & 7.1 & 0.0166 & (0.53,0.42) & 0.87  \\

1/4 & 2.195 & 1.22 & 0.53 &  946 & 5.2 & 0.0194 & (0.46,0.44) & 0.98  \\

\hline
\end{tabular}
\label{tab:phys_props}
\end{center}
\end{table*}

We have found that the internal properties of haloes are robust in the
sense that the concentrations and widths of their distributions and the values
of their medians are practically unaffected as the ``effective'' box size is varied. 
However, we have also found that properties such as the triaxiality 
$T$ and spin $\lambda'$ of individual haloes can vary by a factor 
of several between those truncated runs and the fiducial run. In this short
case study, we illustrate how changing the effective box size impacts on
an individual halo. 

\begin{figure}
  \centering
  \includegraphics[width=8cm]{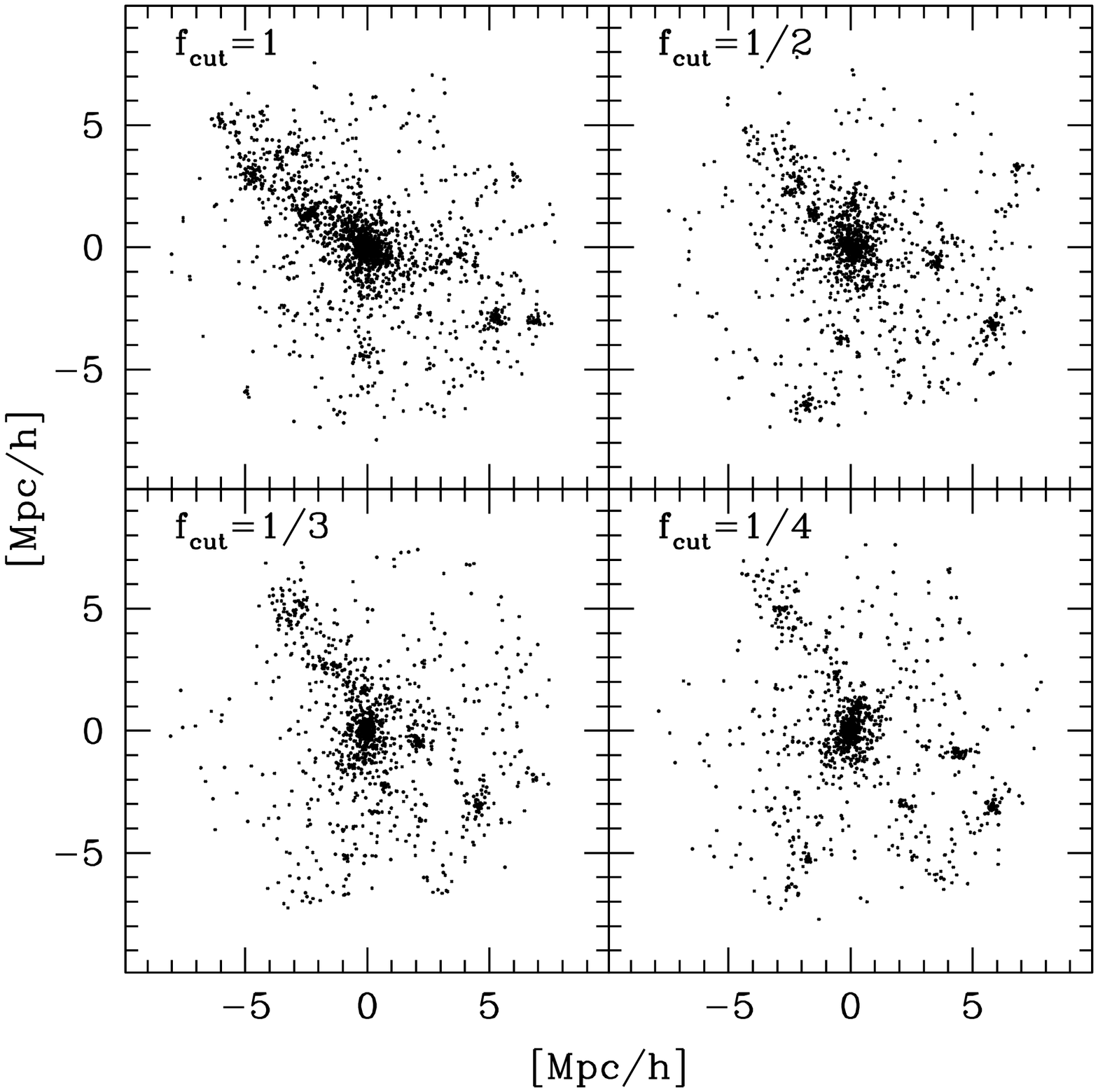}
  \caption{Projected particle distribution in the neighbourhood of a
           $10^{14} h^{-1}\,{\rm M_{\odot}}$ cluster mass halo in each
	   of the four runs. The virial mass of the halo decreases as
	   the power spectrum is truncated at progressively shorter wavelengths.}
\label{fig:haloes_vradial}
\end{figure}

We considered a number of characteristic internal properties of a single well 
resolved halo that we have identified in each of the runs; these are
summarised in table~\ref{tab:phys_props}, while projections of the particle
distribution within 1.5 times the virial radius are shown in 
figure~\ref{fig:haloes_vradial}. We have chosen the most massive halo
in the fiducial simulation for our analysis because we expect the 
differences to be most pronounced on this mass scale.

Inspection of table~\ref{tab:phys_props} reveals that truncation of the
power spectrum has a dramatic effect on the halo's mass, as expected
from figure~\ref{fig:MassMass} -- a factor of $\sim 2$ between the
fiducial run and the three truncated runs. This is
reflected in the decrease in concentration (by $30-40\%$). Interestingly,
the spin parameter increases steadily by up to a factor of $\sim 2$ between
the fiducial and the \texttt{Chop32} truncated run. Also, we note that the
semi-minor axis $c/a$ remains fixed (at $\sim 0.43$) but the semi-major axis
steadily decreases (from 0.73 to 0.46), leading to an increase in triaxiality 
(from T=0.57 to T=0.98).

\section{Discussion}
\label{sec:discussion}

The results presented in the previous show that the size of simulation
box has little impact, if any, on the internal structure and kinematics
of simulated Cold Dark Matter haloes. At first, this result is a little 
surprising, for reasons given in \S~\ref{sec:intro}. Although we focused 
on concentration, shape and spin parameter because we expect these quantities  
to be most susceptible to the contribution of long wavelength perturbations,
we also examined a number of other quantities, including radial velocities, 
velocity anisotropies and the virial ratio $2T/|W|$. In all cases, the 
distributions recovered from the truncated runs are indistinguishable from
those derived from the fiducial run.

However, it can be argued that this result is not so surprising. The impact 
of large scale perturbations on the value of $\sigma_8$ noted by GB94 is 
certainly noticeable -- of order a $40\%$ reduction for a box size of $L_{\rm
  box} = 32 h^{-1}\,{\rm Mpc}$ relative to the ``true'' value -- but the 
effect is less pronounced as we probe smaller mass scales. For example, 
the difference is less than $\sim 10\%$ at $2 h^{-1}\,{\rm Mpc}$ for the same
box size. In other words, large scale perturbations become less important as
we probe smaller masses. We can understand this in figure~\ref{fig:MassMass}, 
which both confirms the result of BR05 that it is the higher mass systems that
 are most affected by the truncation of long wavelength perturbations and also
 highlights that the typical or median halo mass below $\lesssim 10^{13} 
h^{-1}\,{\rm M_{\odot}}$ is unaffected by truncation. 

Perhaps more pertinently, the existence of a universal mass profile 
\citep[][]{nfw96,nfw97} that provides an adequate characterisation of haloes
in a range of hierarchical cosmologies implies that large scale perturbations 
have little effect on the inner structure of dark matter haloes. Indeed,
truncating the power spectrum at short wavelengths produces a halo mass 
profile that is well described by the \citet{nfw96,nfw97} form 
\citep[][]{moore99,Knebe02}. Nevertheless, our results show that the internal
structure of haloes (in the spherically averaged sense) are unaffected by
the introduction of a long wavelength cut-off in the power spectrum \citep[see also][]{tormen96,cole97}.

Similarly, the weakness of the trend in the distribution of halo shapes most 
likely reflects the relevance of the dependence of shape on merging
history. While the infall of material on the halo is sensitive to the
global tidal field, and therefore long wavelength perturbations, a
change in the tidal field does not necessarily suppress infall; rather
it will change the time at which it occurs and therefore will not
affect halo shape in an average sense.

The small effect noted in the spin distributions -- an offset in the median 
values of the truncated and fiducial values -- suggests that long wavelength
perturbations play some role in shaping the spin and consequently the angular
momentum content of the halo. Tidal torques make an important contribution to
the acquisition of angular momentum of the halo during its growth in the linear
regime \citep[][]{white84} but we do not expect the halo's subsequent 
non-linear evolution to depend on the large scale tidal field. The small 
difference likely represents an imprint of the angular momentum acquired 
during the haloes linear growth.

\section{Conclusions}
\label{sec:conclusions}

The aim of this study has been to establish whether finite box size has a
measurable effect on the internal properties of simulated Cold Dark Matter 
haloes. The results of our analysis suggest that the effects, if present, are
small and not statistically significant in most cases. Of the principal 
quantities we have examined -- concentration, shape and the spin parameter -- 
we find that spin shows the most prominent effect; the median spin
parameter is $\sim 50\%$ 
smaller in our truncated runs, independent of the mass of the system. We argue
 that this is an imprint of the linear growth phase of the halo's angular 
momentum by tidal torquing, and that the absence of any measurable trend in 
concentration or strong trend in shape reflect the importance of
virialisation and complex mass accretion histories for these quantities 
respectively.

These results are useful because they clarify what properties of simulated 
dark matter haloes are affected by finite box size, and the severity of these
effects. Indeed, they are of some importance because they reveal that studies 
of the internal properties of statistical samples of haloes that do not 
require clustering information that are based on high resolution simulations 
of small volumes are not compromised. The abundance of high mass systems will 
be underestimated, but the properties of low- to intermediate-mass systems are 
reliable. This is of interest to studies of, for example, the formation of the
 first generation of population III stars at high redshift 
\citep[e.g.][]{abel02}; in these cases, clustering information is not as 
important (although it would be for reionisation studies, say).

In conclusion, our study has demonstrated that the effects of finite box size 
are negligible for the internal properties of dark matter haloes.

\section*{Acknowledgements}
The simulations presented in this paper were carried out on the
Beowulf cluster at the Centre for Astrophysics~\& Supercomputing,
Swinburne University. The financial support of the Australian Research
Council is gratefully acknowledged.

\label{lastpage}
\end{document}